\documentclass{appolb}
\usepackage{epsfig}
\usepackage{slashed}

\begin{document}
\title{Transport Properties of Strong-Interaction Matter%
\thanks{Presented at the HIC for FAIR Workshop and XXVIII Max Born Symposium
{\it Three days on Quarkyonic Island}, Wroc\l aw, May 19 - 21, 2011.}%
}
\author{J. Wambach, K. Heckmann and M. Buballa
\address{Institut f\"ur Kernphysik, 
  Technische Universit\"at Darmstadt,
  D-64289 Darmstadt, GERMANY}
}
\maketitle
\begin{abstract}
The properties of strong-interaction matter are probed in ultra-relativi\-stic heavy-ion collisions. In the context of measurements of the elliptic flow at RHIC and the LHC the shear viscosity is of particular interest. In this presentation we discuss recent results for $\eta/s$ in hadronic matter at vanishing baryo-chemical potential within kinetic theory. Using the Nambu Jona-Lasinio model, special attention is paid to effects arising from the restoration of spontaneously broken chiral symmetry with increasing temperature.  
\end{abstract}
  
\section{Introduction}

\begin{figure}[h]
\begin{center}
\parbox{.58\textwidth}{\vspace{.1\textheight}
\includegraphics[height=.24\textheight]{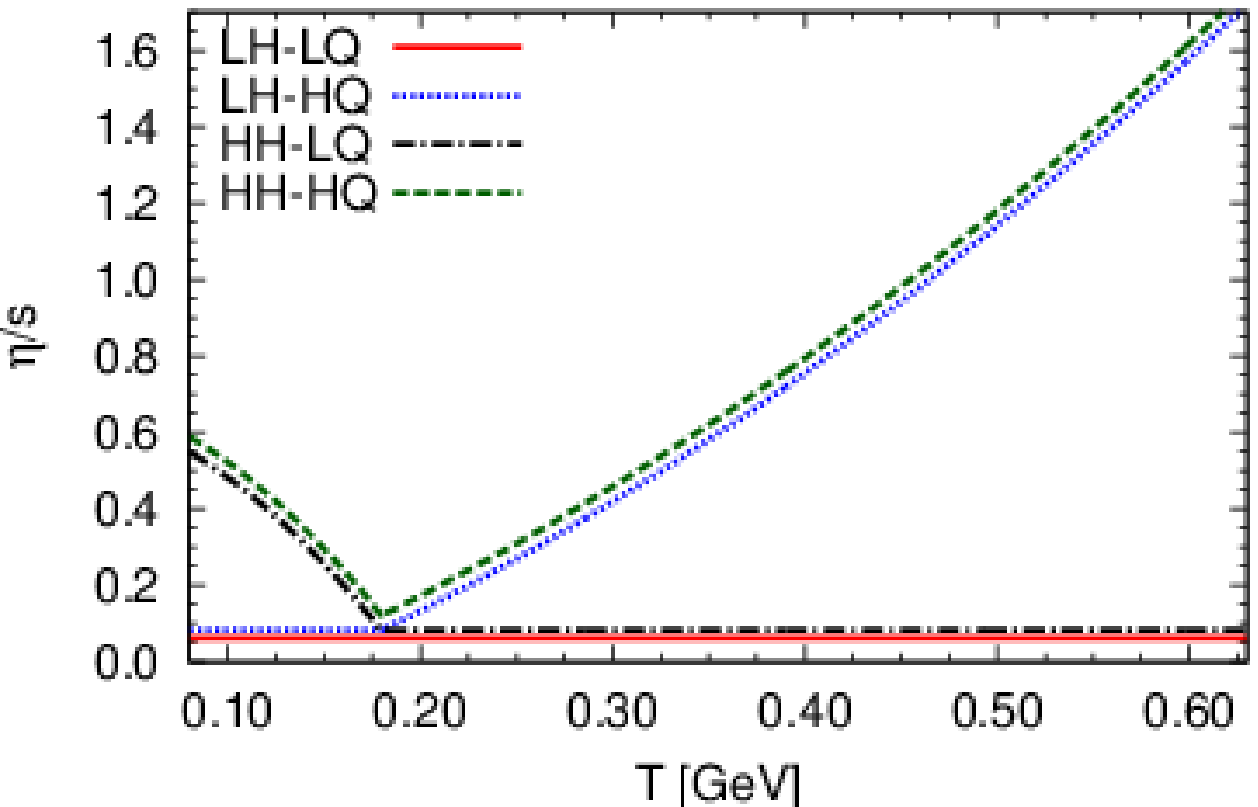}\vspace{.1\textheight}}
\parbox{.41\textwidth}{
\includegraphics[height=.24\textheight]{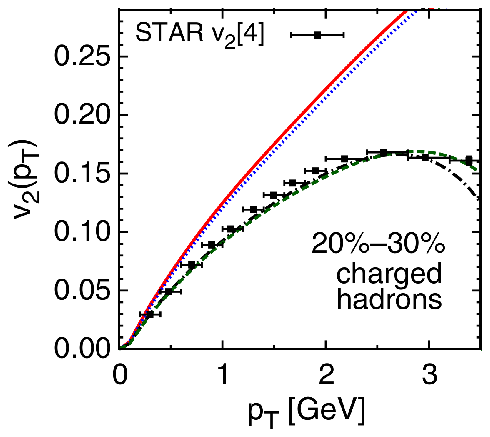} }
  \caption{Assumed temperature dependence of $\eta/s$ (left panel) and the resulting hydrodynamical predictions for $v_2(p_T)$ and their comparison to the STAR data (right panel) \cite{Niemi}.}
  \end{center}
 \label{Fig.Niemi}
\end{figure}

In the exploration of the phase diagram of strong-interaction matter with heavy ions dissipative effects continue to attract large attention. When describing the fireball evolution in viscous hydrodynamics and comparing it  to elliptic flow data from RHIC, a very small ratio of shear viscosity to entropy-density, $\eta/s$, has been inferred \cite{Adare,Adams,Romatschke1}. Similar results have been obtained recently from the ALICE measurements at the LHC \cite{Niemi}. These observations led to the conclusion that strong-interaction matter in the vicinity of the quark-hadron transition behaves almost like a ``perfect fluid''. Recent hydrodynamical calculations of the elliptic flow \cite{Niemi} indicate that the $p_T$-dependence of the flow parameter $v_2$ at RHIC and LHC energies is rather insensitive to the viscous properties of the quark-gluon plasma (QGP) phase and the hadronic phase seems to be more important (Fig.\ref{Fig.Niemi}). 
This calls for detailed understanding of the temperature evolution of $\eta/s$. 

At very low temperatures the physics is dominated by a dilute pion gas whose transport properties are governed by kinetic theory. The collision term is uniquely specified by the current algebra results for the pion-pion scattering length. As the temperature reaches the chiral restoration transition, however,  strong medium modifications of the $\pi\pi$-scattering amplitude are expected due to the softening of the chiral $\sigma$-mode. For a quantitative evaluation of these effects QCD-like theories such as the Nambu Jona-Lasinio (NJL) model are well suited. 
A particularly interesting feature of this model is the fact that mesons
do not exist as elementary degrees of freedom but emerge as composite 
as quark-antiquarks states. As a consequence, the pion, which is a bound state 
below the chiral restoration temperature, becomes a broad resonance at high
temperatures. 
When constructing the scattering amplitude for these objects,
a controlled approximation scheme has to be used, which is consistent with spontaneously broken chiral symmetry and the resulting Goldstone theorem. It is well known that such a scheme is provided by the $1/N$-expansion of the effective action,
where in our case $N$ can be identified with the number of colors $N_c$.

\section{The QCD Phase diagram}

\begin{figure}[h]
\begin{center}
\hspace{-1.0cm} \includegraphics[height=.4\textheight]{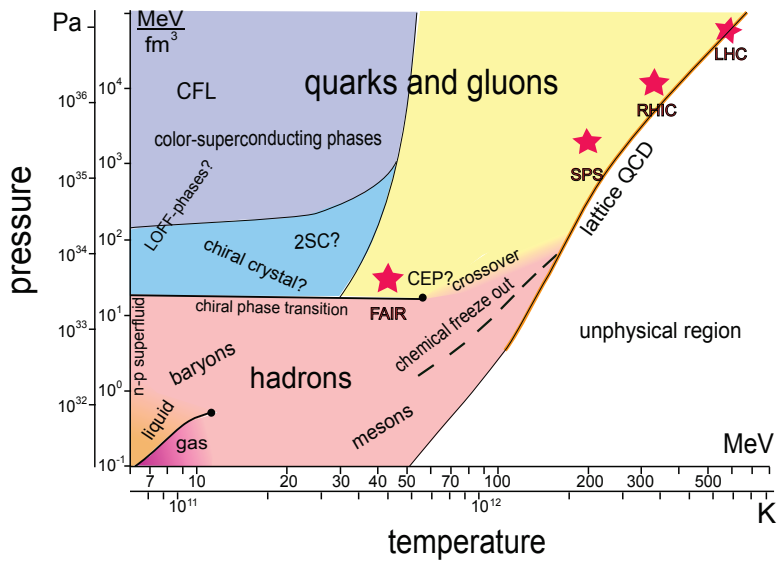} 
\caption{Phase diagram of strong-interaction matter in the pressure-temperature plane \cite{Heckmann1}. Due to relativistic effects there exists an unphysical region in which QCD matter 
cannot exist in equilibrium. }
\end{center}
\label{Fig.PD}
\end{figure}

Before discussing calculations for the $\eta/s$ ratio for confined matter, let us present a novel form of displaying the phase diagram of QCD matter, \ie matter, where the mean interparticle spacing is of the order of a few femtometers. In this case the strong interaction is the main player in the equation of state. Rather than representing the phase diagram in terms of temperature $T$ and baryo-chemical potential $\mu$ we choose to plot pressure vs.\@ temperature. This has the advantage of a more direct comparison with other substances such as water or liquid Helium. The results are shown in Fig.~\ref{Fig.PD}.

\noindent
The low-temperature regime is the realm of nucleonic matter, which may undergo a 
first-order chiral restoration transition to chirally ordered and superconducting quark matter at high pressure. These phases could be realized in the interior of neutron stars. At high temperatures one encounters quark-gluon matter, whose boundary to the unphysical region ($\mu=0$) is quantitatively described by lattice QCD and a free pion gas at low $T$. When raising the temperature the first-order chiral transition line ends in a chiral critical endpoint (CEP) of second order. Current and future heavy-ion experiments are indicated as well as the chemical freeze out. The latter has been accurately determined from particle ratios using the hadron resonance gas model, for which the pressure can be readily obtained from fits of the freeze-out temperature and chemical potential \cite{PBM}.

\section{Shear Viscosity of the hadronic phase}

\begin{figure}[h]
\begin{center}
\includegraphics[width=.23\textheight,angle=-90]{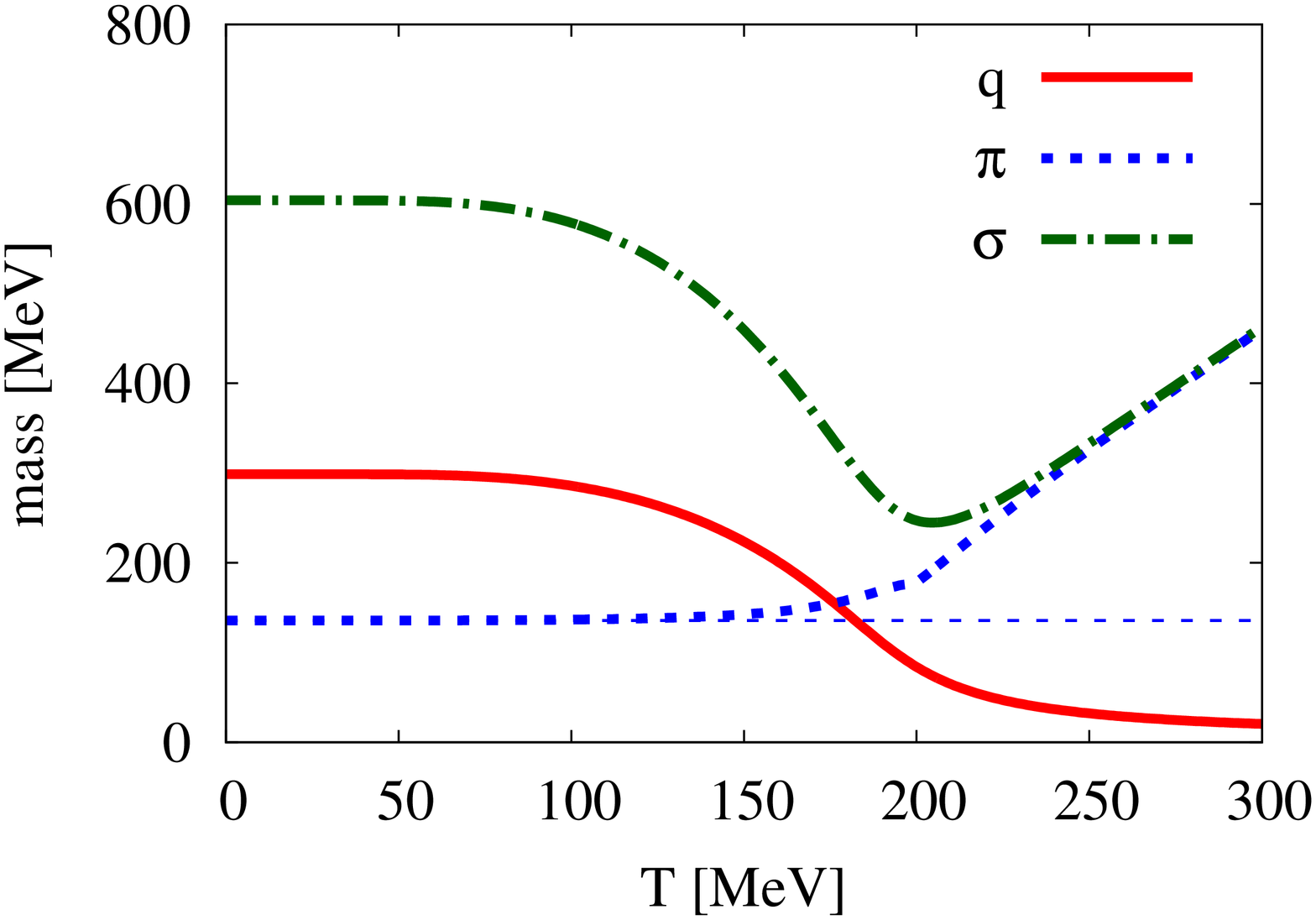} 
\includegraphics[width=.23\textheight,angle=-90]{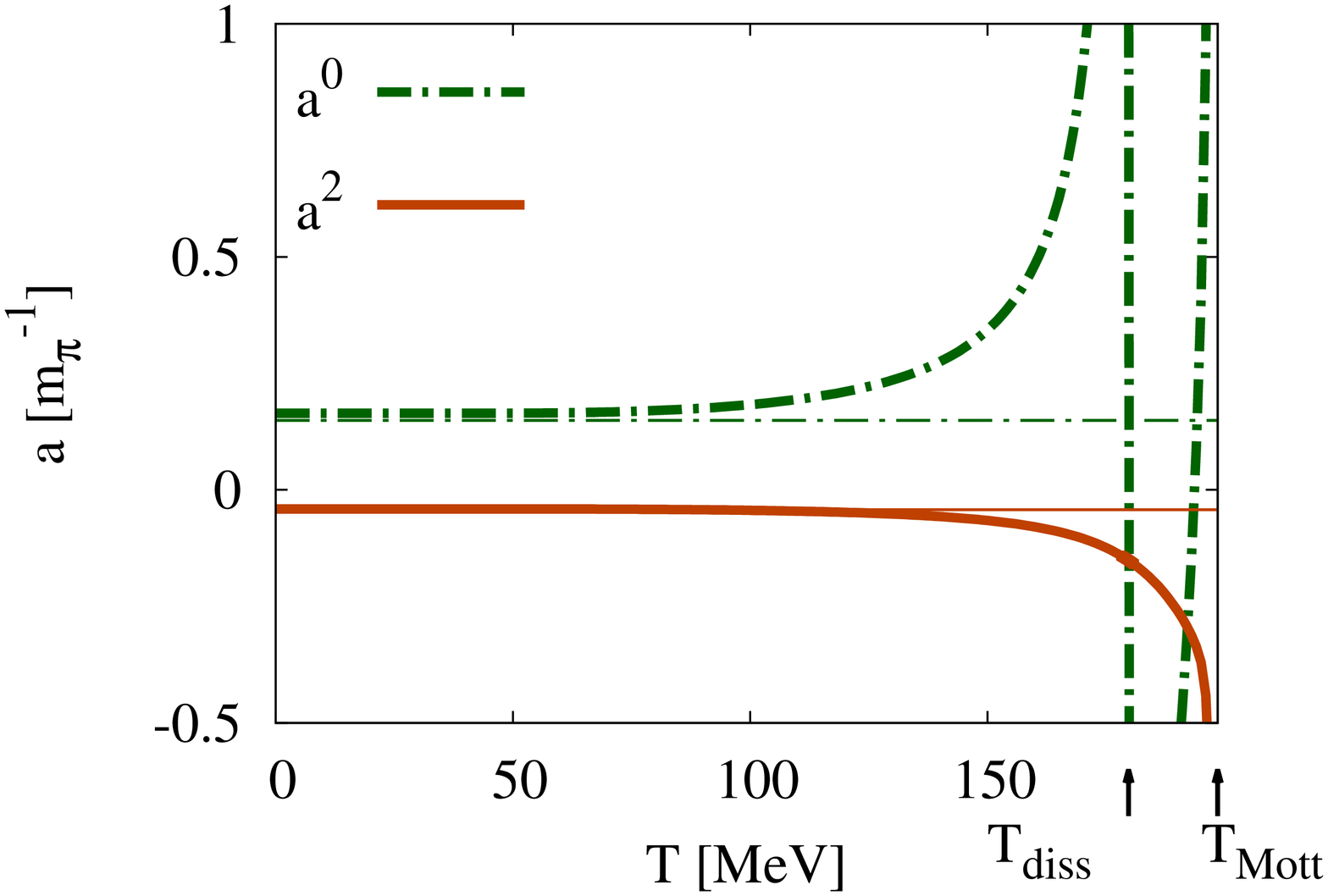} 
\caption{Quark and meson masses (left panel) and scattering lengths 
(right panel) in the two-flavor NJL model as functions of temperature.
The thin lines in the right panel indicate the Weinberg values 
of the vacuum scattering lengths, Eq.~(\ref{eq:aW}).}
  \end{center}
 \label{Fig.Masses}
\end{figure}

It is well known that the low-temperature behavior of the shear viscosity can be described in a Boltzmann-Ueling-Uhlenbeck (BUU) approach in which the time evolution of the one-body phase-space density $f_i$ of a quantum particle, in our case a pion, is given by
\begin{equation}
\frac{\mathrm{d}}{\mathrm{d} t} f_\pi(\vec x,\vec p,t)=C_{\pi\pi}\left [f_\pi\right ]\, ,
\end{equation} 
where the two-body collision integral $C_{\pi\pi}$ includes the transition matrix 
${\cal M}_{\pi\pi}$ for $\pi\pi$-scattering  and the phase-space occupancy in the usual way.     
In the Chapman-Enskog expansion \cite{Chapman} for $f_\pi$ the resulting expression for $\eta$ in a pion gas 
reads (see \eg Refs.~\cite{DLlE:04,IOM:08})
\begin{equation}
 \eta=\frac{1}{5}\frac{4\pi}{(2\pi)^3}\int_0^\infty\!\!\!\!\! dp\frac{p^4}{E_p}
f^{(0)}_\pi(1+f^{(0)}_\pi){\cal B}_\pi(p) \,,
\end{equation}
where $f^{(0)}_\pi$ is the equilibrium distribution and ${\cal B}_\pi(p)$ contains ${\cal M}_{\pi\pi}$. 
When evaluating $\pi\pi$-scattering in the two-flavor NJL model with the Lagrange density  \cite{NJL} 
\begin{equation}
    {\cal L}
    =
    \overline{q}(i \slashed{\partial} -m_0)q
    +
    g[(\overline{q}q)^2+(\overline{q}i\gamma_5\vec{\tau}q)^2],
\end{equation}
the lowest nontrivial order is given by the following two 
diagrams ~\cite{Schulze,Quack}:

\begin{equation}
i{\cal{M}}_{\pi \pi} = 
        \parbox{6ex}{\includegraphics[height=7ex]{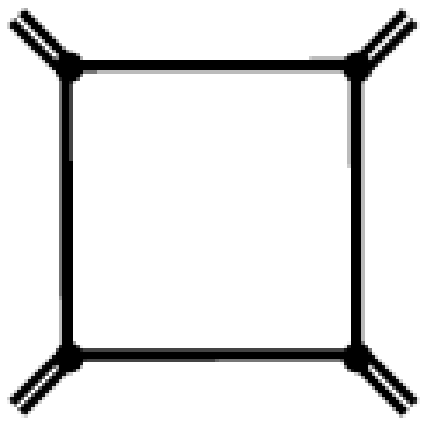}} \quad + \quad
        \parbox{6ex}{\includegraphics[height=7ex]{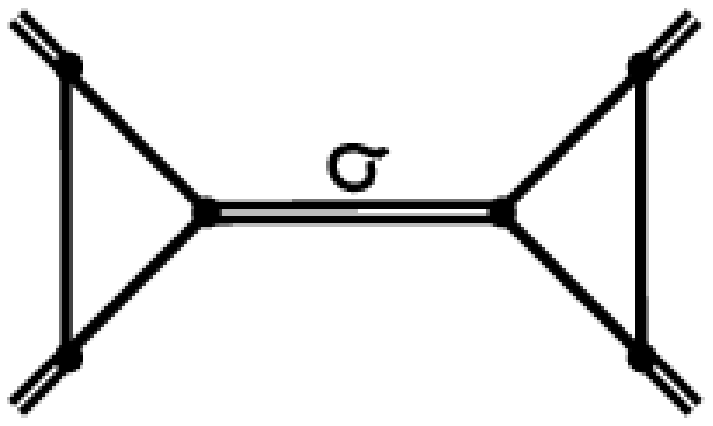}}  
        \hspace{10mm} , 
\label{pipidiagrams}
\end{equation}
where the full lines represent Hartree quarks, 
\begin{equation}
        \parbox{40ex}{\includegraphics[height=10ex]{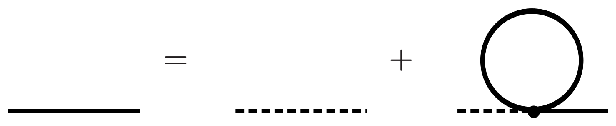}}\quad , 
\end{equation}
while the double lines denote mesons in the Random-Phase approximation,
\begin{equation}
        \parbox{40ex}{\includegraphics[height=9ex]{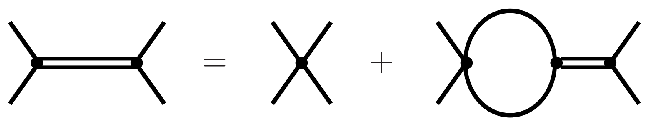}}\quad . 
\end{equation}
The resulting ``in-medium'' quark and meson masses are shown in the left
panel of Fig.~\ref{Fig.Masses}.
We have also indicated two characteristic temperatures~\cite{Quack},
which will be important for the later discussion: 
the ``dissociation temperature'', $T_\mathit{diss}$, where the 
$\sigma$-meson mass becomes degenerate with the in-medium two-pion threshold 
as well as at the ``Mott temperature'', $T_\mathit{Mott}$, where the pion 
dissolves into a quark-antiquark pair.

\subsection{Scattering length}

The scattering lengths in the isospin $I$ channel
are related to the scattering matrix at threshold as
\begin{equation}
a^I = \frac{1}{32 \pi m_\pi}{\cal{M}}_{\pi \pi}^I(s=4m_\pi^2).  
\label{eq:aI}
\end{equation}
To leading order in the pion mass
their vacuum values are entirely dictated by chiral symmetry 
and have been calculated by Weinberg in the 1960s \cite{Weinberg},  
\begin{equation}
a^0_W=\frac{7 m_\pi}{32\pi f_\pi^2},\qquad 
a^2_W=-\frac{2 m_\pi}{32\pi f_\pi^2},   
\label{eq:aW}
\end{equation}
while the isospin-1 scattering length vanishes because of the total
symmetry of the bosonic wave function.

These values are well reproduced by the NJL model when the diagrams
(\ref{pipidiagrams}) are evaluated in vacuum~\cite{Schulze, Quack}.
However, as shown in the right panel of Fig.~\ref{Fig.Masses}, 
with increasing temperature there are important medium 
modifications~\cite{Quack}.
In particular the softening of the $s$-channel $\sigma$-meson
leads to a sharp peak of $a^0$ at $T_\mathit{diss}$, which is
reminiscent to the physics of a Feshbach resonance in ultra-cold atomic gases. 
Similarly $a^0$ and $a^2$ diverge at $T_\mathit{Mott}$
due to threshold singularities of the quark triangle and box diagrams.

\subsection{Shear viscosity}

\begin{figure}[h]
\parbox{.45\textwidth}{
\includegraphics[width=.23\textheight,angle=-90]{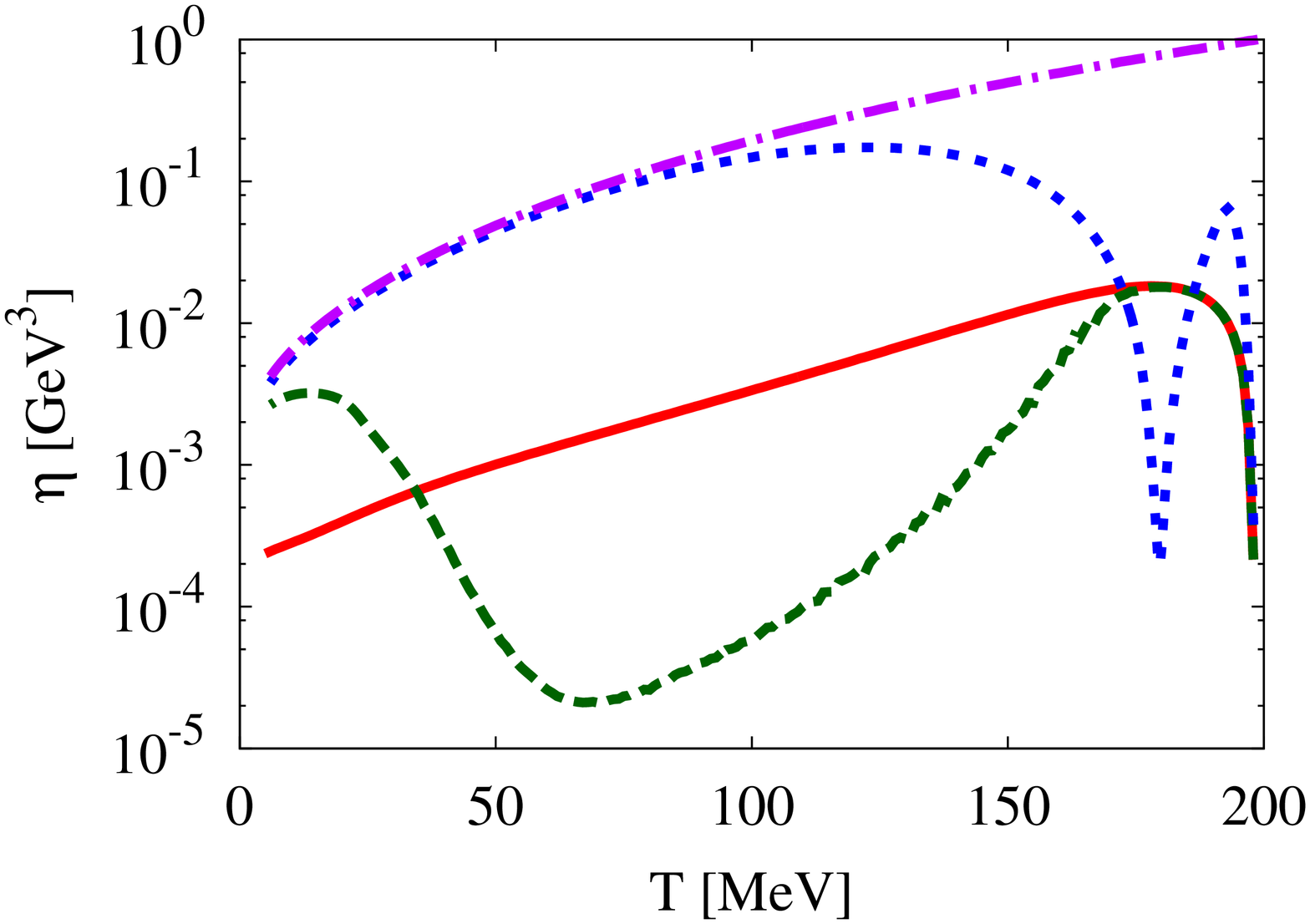}  
}
\hspace{2mm}
\parbox{.45\textwidth}{
\includegraphics[height=.27\textheight]{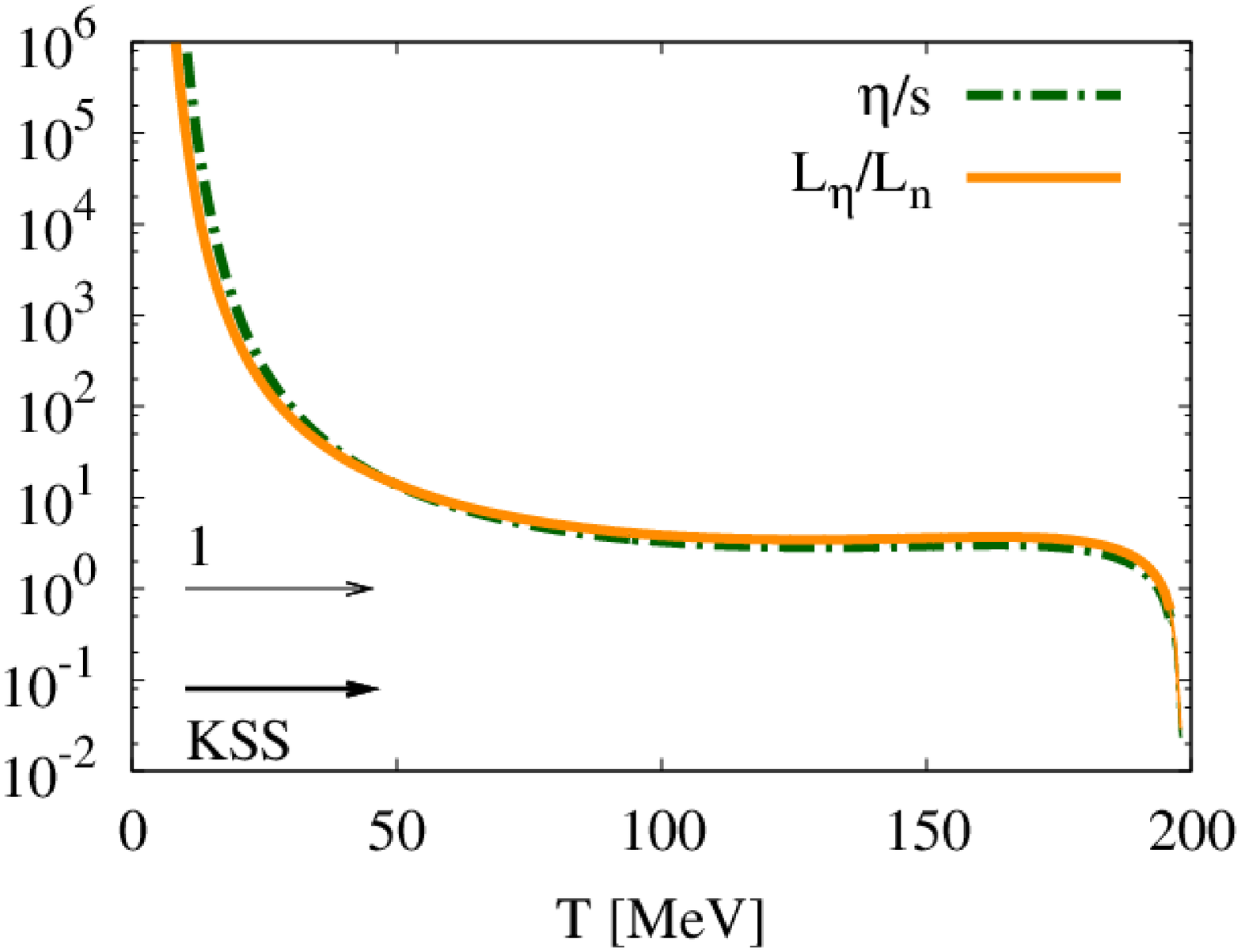}
}
\caption{Left panel: shear viscosity $\eta$ as a function of temperature for various approximations of the $\pi\pi$ cross section \cite{Heckmann}. The dash-dotted line displays the result obtained by using the Weinberg value for the $\pi\pi$-scattering lengths, while for the dotted line medium effects of the scattering lengths are included. The dashed line shows results with the inclusion of the$\sqrt{s}$-dependence of the intermediate $\sigma$-meson propagator.
The full line includes the coupling of the $s$-channel $\sigma$ meson to two-pion states. 
Right panel: $\eta/s$ and the fluidity measure of Ref. \cite{LiaoKoch} for
our most realistic approximation (solid line of the left panel).
The AdS/CFT bound (KSS) \cite{KSS} is also indicated. }
 \label{Fig.Coeff}
\end{figure}

In the left panel of Fig.~\ref{Fig.Coeff} the shear viscosity is displayed as
a function of temperature, using different approximations for the 
scattering amplitude. 
In the simplest case (dash-dotted line), both, momentum and temperature 
dependence are neglected, and ${\cal{M}}_{\pi \pi}^I$ is obtained from 
Eq.~(\ref{eq:aI}) employing the Weinberg values (\ref{eq:aW}) for the 
scattering lengths. 
When the latter are replaced by the temperature dependent NJL-model results,
we obtain the viscosity indicated by the dotted line. 
Whereas at low temperature it is in good agreement with the Weinberg result,
$\eta$ becomes very small in the vicinities of $T_\mathit{diss}$ and 
$T_\mathit{Mott}$, because of the large scattering lengths.
In these regions, however, kinetic theory breaks down, since the mean 
interparticle spacing is no longer large compared to the range of the 
interaction.   

Moreover, because of thermal motion, the approximation of the momentum 
dependent scattering amplitude by its value at threshold becomes 
inappropriate when the temperature increases.
In particular, the pole of the $s$-channel $\sigma$-meson exchange 
(second diagram in (\ref{pipidiagrams})) can be reached 
at temperatures well below the dissociation temperature,
whereas at $T = T_\mathit{diss}$, when this pole is at threshold, most pion 
pairs have much higher momenta. As a consequence, the sharp minimum at 
$T = T_\mathit{diss}$ gets washed out and shifted to lower temperatures
when the momentum dependence of the $\sigma$-propagator
is taken into account (dashed line). For simplicity, we still neglect the 
momentum dependence of the quark triangles and boxes in our calculations. 
Therefore, the steep drop of the viscosity near the Mott temperature remains. 

Another not very realistic feature of this approximation is the fact that
the $\sigma$-meson is a sharp resonance in Random-Phase approximation.
This can be remedied by including a $\sigma$ width from two-pion decay in the 
$K$-matrix approximation. One can show that this still fulfills the dilute gas limit, given by the vacuum scattering lengths. The resulting shear viscosity 
is given by the full line, which should be considered as the most reliable result of the present study. 

The corresponding values for $\eta/s$ and the fluidity measure $L_\eta/L_n=\eta n^{1/3}/h c_s$ \cite{LiaoKoch}, where $h$ denotes the enthalpy and $c_s$ the speed of sound, are displayed 
in the right panel of Fig. \ref{Fig.Coeff} as the dash-dotted and full line,
respectively. For the entropy density $s$, the particle density $n$, and the
speed of sound $c_s$ we took ideal-gas values, which is consistent with
the dilute-gas assumption of the BUU approach.
It turns out that both fluidity measures are very similar. 
After decreasing by several orders of magnitude at low temperatures, 
the curves become rather flat in an intermediate temperature regime, 
where we find $\eta/s \approx 3$.  
As already mentioned, the steep drop near the Mott temperature is most
likely an artifact of neglecting the momentum dependence of the quark
triangles and boxes.

\section{Discussion}

The in-medium $\pi\pi$ cross section has been evaluated in the two-flavor NJL 
model with the aim to include effects of chiral restoration with increasing 
temperature.  
This leads to important modifications of $\eta/s$ at finite temperature, 
which render the viscous effects much smaller than it would be expected from 
a simple vacuum extrapolation. 
On the other hand, the results turn out to be extremely sensitive to the
applied approximations, and our ``most reliable'' model is certainly not
the last word.
In fact, except for the region close to the Mott temperature, which should
not be trusted, our results are still more than one order of magnitude
above the KSS bound $\eta/s = 1/4\pi$~\cite{KSS}.
Various improvements and extensions of the model should therefore be
performed:

In the $\pi\pi$ sector, we should include intermediate $\rho$ 
mesons, in order to get a realistic description of the $p$-wave isovector 
channel. 
We should also include the scattering of other particles, which are suppressed
at low temperatures, but can become important in the crossover region. 
In particular we wish to extend the model to three flavors and include
kaons and $\eta$-mesons.
It would also be interesting to include the scattering of quarks, which 
should become important above the crossover temperature.
Work in these directions is in progress.
\vspace{1cm}

\noindent
{\bf Acknowledgement}
\vspace{0.5 cm}

\noindent
This work was supported in part by the Helmholtz International Center for FAIR, the 
Helmholtz Institute EMMI and the BMBF grant 06DA9047I.

\end{document}